\documentclass[10pt,letterpaper]{article}
\usepackage[hidelinks]{hyperref}

\usepackage{ccn}
\usepackage{pslatex}
\usepackage{apacite}
\usepackage{graphicx}

\usepackage{booktabs}
\usepackage{multicol}
\usepackage{multirow}
\usepackage{setspace}
\usepackage{tikz}
\usepackage{caption}

\title{Modeling infant object perception as program induction}

\author{\\{\large \bf J.-Philipp Fränken$^{1*}$, Christopher G. Lucas$^2$, Neil R. Bramley$^2$, Steven T. Piantadosi$^3$}\\
     $^1$Stanford University, $^2$The University of Edinburgh,  $^3$University of California, Berkeley, 
     $^*$jphilipp@stanford.edu\\}

\begin{document}

\maketitle

\section{Abstract}
{
\bf
Infants expect physical objects to be rigid and persist through space and time and in spite of occlusion. 
Developmentists frequently attribute these expectations to a ``core system'' for object recognition.
However, it is unclear if this move is necessary. If object representations emerge reliably from general inductive learning mechanisms exposed to small amounts of environment data, it could be that infants simply induce these assumptions very early. 
Here, we demonstrate that a domain general learning system, previously used to model concept learning and language learning, can also induce models of these distinctive ``core'' properties of objects after exposure to a small number of examples. 
Across eight micro-worlds inspired by experiments from the developmental literature, our model generates concepts that capture core object properties, including rigidity and object persistence. 
Our findings suggest infant object perception may rely on a general cognitive process that creates models to maximize the likelihood of observations.}\footnote{\tiny Project page: \href{https://janphilippfranken.github.io/object-perception/}{\texttt{\tiny janphilippfranken.github.io/object-perception}}}

\begin{quote}
\small
\textbf{Keywords:} 
core knowledge; perception; vector quantization; program induction; Bayes
\end{quote}
\vspace{-.3cm}
\section{Introduction}

Object representations serve as compositional building blocks for higher level cognition in both humans and machines \cite{xu1996infants, scholkopf2021toward, chen2022unsupervised}. Developmental accounts suggest that infants rely on a ``core system'' for object representations to perceive the boundaries of objects, accurately represent their shapes even when they are partially or fully occluded, and make predictions about object movements and their final positions \cite{spelke2007core}. Having a specific system for representing objects from an early age can be beneficial because it allows for the incorporation of prior knowledge and expectations about objects and their physical regularities, such as the idea that objects usually maintain their shape and size as they move (\textit{rigidity} principle; \citeNP{spelke1990principles}) and continue to exist and retain their properties even when occluded (\textit{object persistence} principle; \citeNP{baillargeon1987object, baillargeon2008innate}). Despite converging evidence for the existence of a core object system in both human infants (e.g., \citeNP{feigenson2003tracking, spelke2022babies}) and non-human animals (e.g., \citeNP{chiandetti2015inexperienced, hauser2003spontaneous}), it is not clear if a system specifically designed for this purpose is necessary or beneficial if object representations can be learned effectively by a domain general inductive system from only a small amount of data.

\begin{figure}[]
  \centering
  \begin{tikzpicture}
    \node at (0,0) {\includegraphics[width=0.4\textwidth]{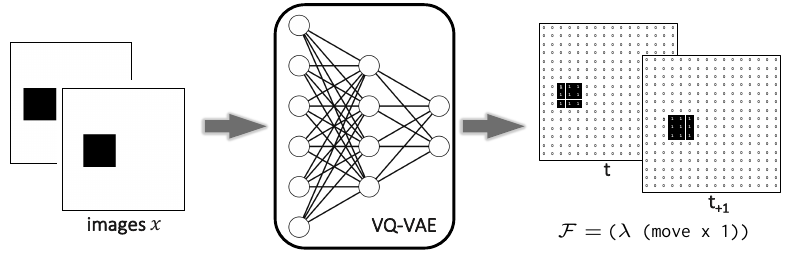}};
\end{tikzpicture}
\vspace{-.5cm}
  \captionsetup{font=small}
  \caption{Illustration of inference pipeline. We take short videos (ten frames) as input. These are preprocessed into a sequence of discrete feature maps using vector quantization followed by one-hot encodings on each feature map to obtain Boolean tensors (or ``bitmasks''). Bitmasks are then processed by a generic Bayesian concept learning algorithm to induce programs that parsimoniously explain the underlying structure in the discretized data. For example, evaluating the program $\mathcal{F}$ will move (``roll'') the upper bitmask ($t$) by \texttt{1} on the \texttt{x} dimension predicting the bitmask shown below ($t_{+1}$).}
  \label{fig:fig_1}
  \vspace{-0.6cm}
\end{figure}

\section{Infant object perception as program induction}
We assume that object representations can be used to efficiently compress and discretize perceptual input (e.g., visual). As such, object representations might arise from reasoning about the physical regularities \cite{spelke1990principles} or ``invariants'' \cite{sloman2004causal} in one's environment that facilitate predictions about its future states. Inspired by recent advances in Bayesian program learning \cite{ellis2021dreamcoder, tang2022perception, yang2022one} and intuitive physics \cite{piloto2022intuitive}, we present an idealized model (Figure~\ref{fig:fig_1}) that discovers object representations and their physical regularities from short sequences of 2D images (which should generalize to 3D scene projections). Our model can be summarized in four steps: (1) Extract a discrete "codebook" representation $c$ for each image $x$ using a VQ-VAE (\citeNP{van2017neural}), a simple tool for efficient image encoding without relying on semantic object assumptions $\rightarrow$ (2) Apply $n$ deterministic one-hot encodings to each discrete feature map $c$ to generate Boolean tensor representations ("bitmasks"), with $n$ representing the number of unique codes in $c$ $\rightarrow$ (3) Use a Bayesian concept learning algorithm to process the resulting bitmasks and generate programs that parsimoniously explain the structure in the data $\rightarrow$ (4) Use discovered programs to improve the representation by searching for structure in residuals or imputing missing data to maximize likelihood. The final two steps are repeated until convergence or until a time-out threshold is reached. To discover programs, our model generates compositions of functions from the primitives listed in Table~1 and computes posterior distributions over programs using Bayes' rule: $P(H \mid D) \propto P(H) P(D \mid H)$. The prior probability of a program $P(H)$ is determined by a probabilistic context-free grammar (PCFG) based on the operations in Table~1. For likelihood $P(D \mid H)$ we assume a standard exponential loss function.
We use stochastic (MCMC) sampling as in \citeA{goodman2008rational} to search for programs.

\begin{table}[t]
\begin{center}
\singlespacing
\footnotesize
\vspace{-.6cm}
\textbf{Table 1.} Assumed primitive functions
\begin{tabular}{@{}l l@{}}
  \toprule
  \textbf{Type} & \textbf{Primitives} \\
  \midrule
  \multirow{2}{*}{Number functions}
  & \texttt{(add n)}, \texttt{(sub n)}, \texttt{(mult n)},\\
  & \texttt{(div n)}, \texttt{(mod n)}, \texttt{(neg)}, \texttt{(const)}\\
  \midrule
  \multirow{1}{*}{Set functions}
  & \texttt{(union)}, \texttt{(intersection)}\\
  \midrule
  \multirow{2}{*}{Bitmask functions}
  & \texttt{(move x n)}, \texttt{(move y n)},\\
  & \texttt{(complement)}, \texttt{(const)}\\
  \bottomrule
\end{tabular}\par
\end{center}
\vspace{-0.1cm}
\footnotesize{
The space of programs consists of all compositions of these functions that respect the input and output types.}
\vspace{-.6cm}
\label{tab:tab_1_primitives}
\end{table}

\begin{figure}[!b]
\vspace{-0.6cm}
\centering
\begin{tikzpicture}
\node at (0,0) {\includegraphics[width=0.46\textwidth]{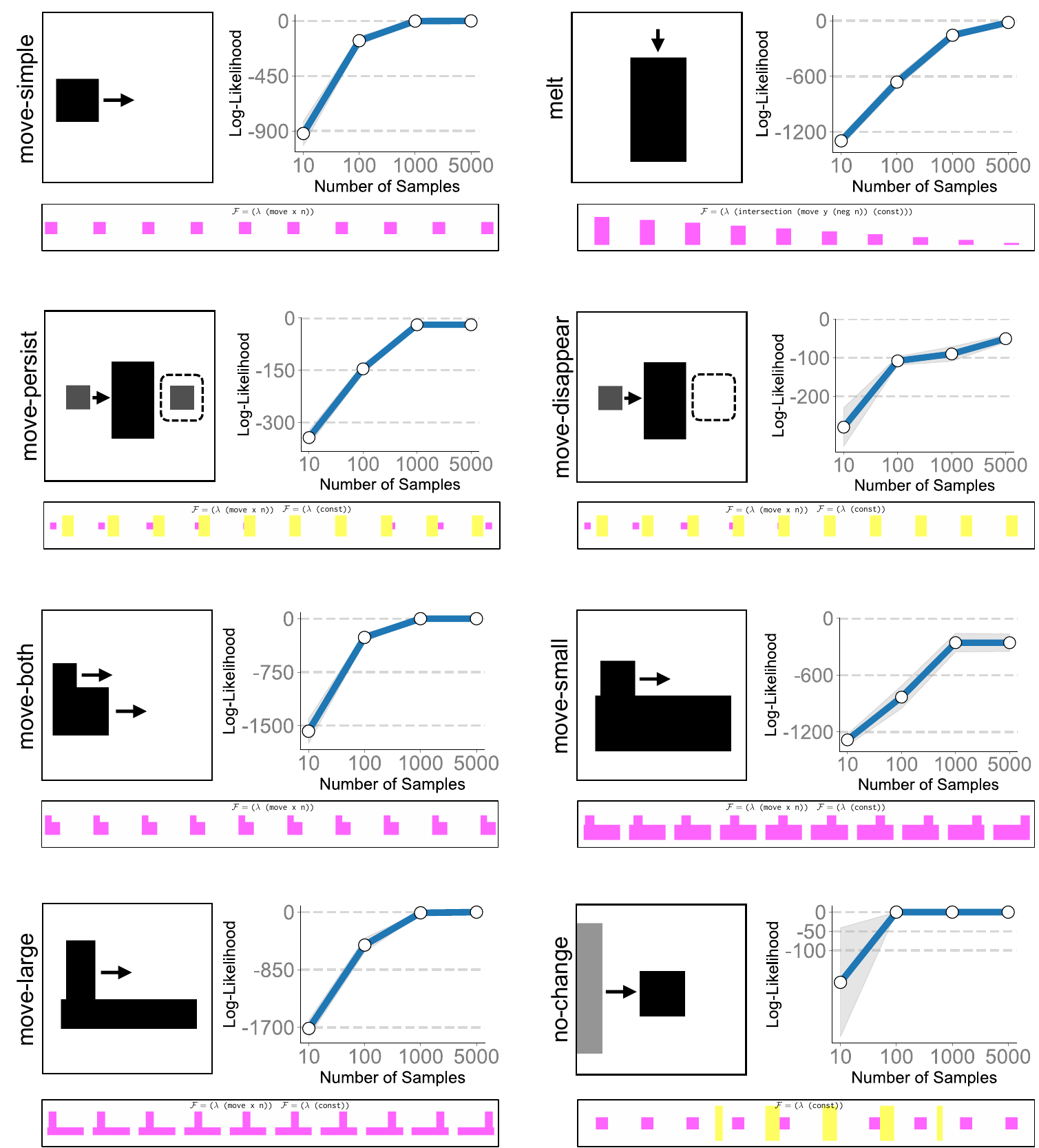}};
\draw (-4,4.75) node[below] {a)}; 
\draw (0.15, 4.75) node[below] {b)}; 

\draw (-4,2.4) node[below] {c)}; 
\draw (0.15, 2.4) node[below] {d)}; 

\draw (-4,0.02) node[below] {e)}; 
\draw (0.15, 0.02) node[below] {f)}; 

\draw (-4,-2.3) node[below] {g)}; 
\draw (0.15,-2.3) node[below] {h)};

\end{tikzpicture}
\vspace{-0.325cm}
\captionsetup{font=small}
\caption{Illustration of tested micro-worlds and learning curves for the model. The $x$ axis corresponds to the number of samples (i.e., the length of the MCMC chain) and the $y$ axis corresponds to the log-likelihood of the final program(s) from a given chain averaged across $100$ independent runs. Grey shadings correspond to standard error. Full sequence for each micro-world and target program(s) are shown at the bottom of each panel.}
\label{fig:fig_2}
\vspace{-0.55cm}
\end{figure}
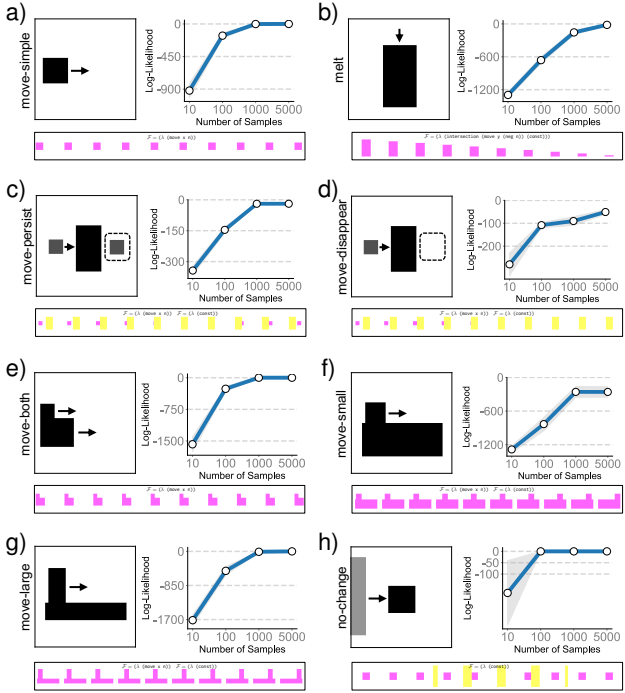

\section{Experiments}
\vspace{-0.05cm}
We evaluate our model's ability to learn object representations and their regularities across eight micro-worlds inspired by experiments from the developmental literature. We use ten images for each probe. To demonstrate our approach, we first examine a baseline probe including simple left-right movement (Fig.~\ref{fig:fig_2}a). To show that we can also handle natural categories that violate standard object properties, we next consider a ``melting'' block (i.e., a block that is shrinking vertically; Fig.~\ref{fig:fig_2}b). We then test the ability of our approach to discover, from sparse input, principles that are often considered as core knowledge, including the widely studied principles of object persistence (\citeNP{baillargeon2008innate, piloto2022intuitive}; Fig.~\ref{fig:fig_2}c--d) and rigidity (\citeNP{spelke1990principles, kemp2008ideal}; Fig.~\ref{fig:fig_2}e--g) . We additionally include an example of unchangeableness following occlusion (\citeNP{baillargeon2012core}; Fig.~\ref{fig:fig_2}h).

\section{Results}
\begin{figure}[!h]
\centering
\vspace{-0.4cm}
\begin{tikzpicture}
\node at (0,0) {\includegraphics[width=0.38\textwidth]{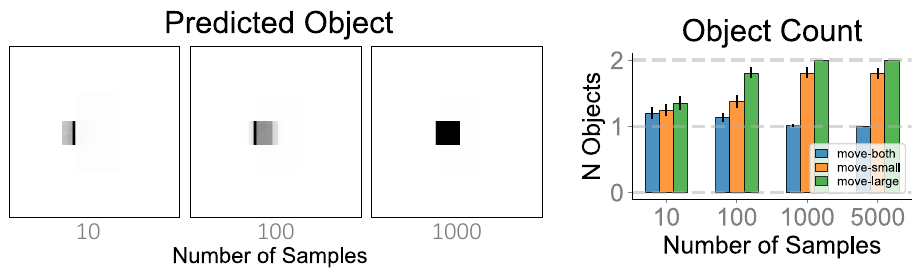}};
\draw (-3.6,1.05) node[below] {a)}; 
\draw (0.95, 1.05) node[below] {b)}; 
\end{tikzpicture}
\vspace{-0.4cm}
\captionsetup{font=small}
\caption{a) Probability of predicted object (greyscale) during occlusion for different numbers of samples. b) Average object counts ($\pm$ SEM) for the three example tests of rigidity.}
\vspace{-0.4cm}
\label{fig:fig_3}
\end{figure}

Panels a--b in Fig.~\ref{fig:fig_2} show that our model can find programs capturing simple object regularities such as constant left-right movement, which can be expressed as \texttt{(@$\lambda$@ (move x n))} as well as ``melting'' which can be expressed as \texttt{(@$\lambda$@ (intersection (move y (neg n)) (const)))}. Figure~\ref{fig:fig_2}c shows that this ability still holds for an object that moves behind an occluder. Figure~\ref{fig:fig_3}a shows the probability of the occluded object for frame $5$ in Figure~\ref{fig:fig_2}c. Consistent with a flattening learning curve at 1000 samples, the model is learning representations at around 1000 samples. The representation of the occluded object was obtained by imputing its representation using a program such as \texttt{(@$\lambda$@ (move x n))} which will have a maximum likelihood if it keeps representing the object during occlusion. 
In line with this idea, the example in Figure~\ref{fig:fig_2}d has a weaker learning curve as the object does not reappear, making it harder to find a physically plausible regularity of the object. Overall, these findings are consistent with increased surprise in infants when objects suddenly disappear or reappear after obstacles as well as their tendency to keep representing objects during occlusion \cite{baillargeon2008innate}. 

Results in Figure~\ref{fig:fig_2}e--g demonstrate our model's ability to interpret ambiguous scenes involving two blocks of different sizes similar to how infants do \cite{kestenbaum1987perception, spelke1989object}. Specifically, our model provides a single-object interpretation for the example shown in Figure~\ref{fig:fig_2}e and a two-object interpretation for the examples shown in Figure~\ref{fig:fig_2}f-g, which require two regularities (both \texttt{(@$\lambda$@ (move x n))} and  \texttt{(@$\lambda$@ (const))}). Average object counts for different numbers of samples are shown in Figure~\ref{fig:fig_3}b, and stable performance is achieved around 1000 sample. Our final example (Figure~\ref{fig:fig_2}h) demonstrates the concept of unchangeableness where an object is occluded by a plank moving across the scene (we do not model the plank's regularity). Following the same imputation approach as in Figure~\ref{fig:fig_2}c, our model can efficiently learn the objects regularity from a small amount of data.

\section{Discussion}
Object representations form a fundamental aspect of human and machine cognition. We proposed that these representations can be learned by domain general learning system that aims to induce symbolic programs to maximize the likelihood of observations. A limitation of the present proof-of-concept results is that we did not jointly train the VQ-VAE and search for programs but instead trained the VQ-VAE prior to search to obtain discrete codebooks. Future work should thus explore joint end-to-end learning of both the VQ-VAE and program search to test our model in more complex scenes (e.g., \citeNP{piloto2022intuitive, maoclevrer}).

\bibliographystyle{apacite}

\setlength{\bibleftmargin}{.125in}
\setlength{\bibindent}{-\bibleftmargin}

\bibliography{ccn}

\begin{thebibliography}{}

\bibitem [\protect \citeauthoryear {%
Baillargeon%
}{%
Baillargeon%
}{%
{\protect \APACyear {1987}}%
}]{%
baillargeon1987object}
\APACinsertmetastar {%
baillargeon1987object}%
\begin{APACrefauthors}%
Baillargeon, R.%
\end{APACrefauthors}%
\unskip\
\newblock
\APACrefYearMonthDay{1987}{}{}.
\newblock
{\BBOQ}\APACrefatitle {Object permanence in 3$1/2$-and 4$1/2$-month-old
  infants.} {Object permanence in 3$1/2$-and 4$1/2$-month-old infants.}{\BBCQ}
\newblock
\APACjournalVolNumPages{Developmental psychology}{23}{5}{655}.
\PrintBackRefs{\CurrentBib}

\bibitem [\protect \citeauthoryear {%
Baillargeon%
}{%
Baillargeon%
}{%
{\protect \APACyear {2008}}%
}]{%
baillargeon2008innate}
\APACinsertmetastar {%
baillargeon2008innate}%
\begin{APACrefauthors}%
Baillargeon, R.%
\end{APACrefauthors}%
\unskip\
\newblock
\APACrefYearMonthDay{2008}{}{}.
\newblock
{\BBOQ}\APACrefatitle {Innate ideas revisited: For a principle of persistence
  in infants' physical reasoning} {Innate ideas revisited: For a principle of
  persistence in infants' physical reasoning}.{\BBCQ}
\newblock
\APACjournalVolNumPages{Perspectives on Psychological Science}{3}{1}{2--13}.
\PrintBackRefs{\CurrentBib}

\bibitem [\protect \citeauthoryear {%
Baillargeon%
\ \BBA {} Carey%
}{%
Baillargeon%
\ \BBA {} Carey%
}{%
{\protect \APACyear {2012}}%
}]{%
baillargeon2012core}
\APACinsertmetastar {%
baillargeon2012core}%
\begin{APACrefauthors}%
Baillargeon, R.%
\BCBT {}\ \BBA {} Carey, S.%
\end{APACrefauthors}%
\unskip\
\newblock
\APACrefYearMonthDay{2012}{}{}.
\newblock
{\BBOQ}\APACrefatitle {Core cognition and beyond: The acquisition of physical
  and numerical knowledge.} {Core cognition and beyond: The acquisition of
  physical and numerical knowledge.}{\BBCQ}
\newblock

\PrintBackRefs{\CurrentBib}

\bibitem [\protect \citeauthoryear {%
Chen%
\ \protect \BOthers {.}}{%
Chen%
\ \protect \BOthers {.}}{%
{\protect \APACyear {2022}}%
}]{%
chen2022unsupervised}
\APACinsertmetastar {%
chen2022unsupervised}%
\begin{APACrefauthors}%
Chen, H.%
, Venkatesh, R.%
, Friedman, Y.%
, Wu, J.%
, Tenenbaum, J\BPBI B.%
, Yamins, D\BPBI L.%
\BCBL {}\ \BBA {} Bear, D\BPBI M.%
\end{APACrefauthors}%
\unskip\
\newblock
\APACrefYearMonthDay{2022}{}{}.
\newblock
{\BBOQ}\APACrefatitle {Unsupervised Segmentation in Real-World Images via
  Spelke Object Inference} {Unsupervised segmentation in real-world images via
  spelke object inference}.{\BBCQ}
\newblock
\APACjournalVolNumPages{arXiv preprint arXiv:2205.08515}{}{}{}.
\PrintBackRefs{\CurrentBib}

\bibitem [\protect \citeauthoryear {%
Chiandetti%
, Spelke%
\BCBL {}\ \BBA {} Vallortigara%
}{%
Chiandetti%
\ \protect \BOthers {.}}{%
{\protect \APACyear {2015}}%
}]{%
chiandetti2015inexperienced}
\APACinsertmetastar {%
chiandetti2015inexperienced}%
\begin{APACrefauthors}%
Chiandetti, C.%
, Spelke, E\BPBI S.%
\BCBL {}\ \BBA {} Vallortigara, G.%
\end{APACrefauthors}%
\unskip\
\newblock
\APACrefYearMonthDay{2015}{}{}.
\newblock
{\BBOQ}\APACrefatitle {Inexperienced newborn chicks use geometry to
  spontaneously reorient to an artificial social partner} {Inexperienced
  newborn chicks use geometry to spontaneously reorient to an artificial social
  partner}.{\BBCQ}
\newblock
\APACjournalVolNumPages{Developmental Science}{18}{6}{972--978}.
\PrintBackRefs{\CurrentBib}

\bibitem [\protect \citeauthoryear {%
Ellis%
\ \protect \BOthers {.}}{%
Ellis%
\ \protect \BOthers {.}}{%
{\protect \APACyear {2021}}%
}]{%
ellis2021dreamcoder}
\APACinsertmetastar {%
ellis2021dreamcoder}%
\begin{APACrefauthors}%
Ellis, K.%
, Wong, C.%
, Nye, M.%
, Sabl{\'e}-Meyer, M.%
, Morales, L.%
, Hewitt, L.%
\BDBL {}Tenenbaum, J\BPBI B.%
\end{APACrefauthors}%
\unskip\
\newblock
\APACrefYearMonthDay{2021}{}{}.
\newblock
{\BBOQ}\APACrefatitle {Dreamcoder: Bootstrapping inductive program synthesis
  with wake-sleep library learning} {Dreamcoder: Bootstrapping inductive
  program synthesis with wake-sleep library learning}.{\BBCQ}
\newblock
\BIn{} \APACrefbtitle {Proceedings of the 42nd ACM SIGPLAN International
  Conference on Programming Language Design and Implementation} {Proceedings of
  the 42nd acm sigplan international conference on programming language design
  and implementation}\ (\BPGS\ 835--850).
\PrintBackRefs{\CurrentBib}

\bibitem [\protect \citeauthoryear {%
Feigenson%
\ \BBA {} Carey%
}{%
Feigenson%
\ \BBA {} Carey%
}{%
{\protect \APACyear {2003}}%
}]{%
feigenson2003tracking}
\APACinsertmetastar {%
feigenson2003tracking}%
\begin{APACrefauthors}%
Feigenson, L.%
\BCBT {}\ \BBA {} Carey, S.%
\end{APACrefauthors}%
\unskip\
\newblock
\APACrefYearMonthDay{2003}{}{}.
\newblock
{\BBOQ}\APACrefatitle {Tracking individuals via object-files: evidence from
  infants’ manual search} {Tracking individuals via object-files: evidence
  from infants’ manual search}.{\BBCQ}
\newblock
\APACjournalVolNumPages{Developmental Science}{6}{5}{568--584}.
\PrintBackRefs{\CurrentBib}

\bibitem [\protect \citeauthoryear {%
Goodman%
, Tenenbaum%
, Feldman%
\BCBL {}\ \BBA {} Griffiths%
}{%
Goodman%
\ \protect \BOthers {.}}{%
{\protect \APACyear {2008}}%
}]{%
goodman2008rational}
\APACinsertmetastar {%
goodman2008rational}%
\begin{APACrefauthors}%
Goodman, N.%
, Tenenbaum, J.%
, Feldman, J.%
\BCBL {}\ \BBA {} Griffiths, T.%
\end{APACrefauthors}%
\unskip\
\newblock
\APACrefYearMonthDay{2008}{}{}.
\newblock
{\BBOQ}\APACrefatitle {{A rational analysis of rule-based concept learning}}
  {{A rational analysis of rule-based concept learning}}.{\BBCQ}
\newblock
\APACjournalVolNumPages{{Cognitive Science}}{32}{1}{108--154}.
\PrintBackRefs{\CurrentBib}

\bibitem [\protect \citeauthoryear {%
Hauser%
\ \BBA {} Carey%
}{%
Hauser%
\ \BBA {} Carey%
}{%
{\protect \APACyear {2003}}%
}]{%
hauser2003spontaneous}
\APACinsertmetastar {%
hauser2003spontaneous}%
\begin{APACrefauthors}%
Hauser, M\BPBI D.%
\BCBT {}\ \BBA {} Carey, S.%
\end{APACrefauthors}%
\unskip\
\newblock
\APACrefYearMonthDay{2003}{}{}.
\newblock
{\BBOQ}\APACrefatitle {Spontaneous representations of small numbers of objects
  by rhesus macaques: Examinations of content and format} {Spontaneous
  representations of small numbers of objects by rhesus macaques: Examinations
  of content and format}.{\BBCQ}
\newblock
\APACjournalVolNumPages{Cognitive Psychology}{47}{4}{367--401}.
\PrintBackRefs{\CurrentBib}

\bibitem [\protect \citeauthoryear {%
Kemp%
\ \BBA {} Xu%
}{%
Kemp%
\ \BBA {} Xu%
}{%
{\protect \APACyear {2008}}%
}]{%
kemp2008ideal}
\APACinsertmetastar {%
kemp2008ideal}%
\begin{APACrefauthors}%
Kemp, C.%
\BCBT {}\ \BBA {} Xu, F.%
\end{APACrefauthors}%
\unskip\
\newblock
\APACrefYearMonthDay{2008}{}{}.
\newblock
{\BBOQ}\APACrefatitle {An ideal observer model of infant object perception} {An
  ideal observer model of infant object perception}.{\BBCQ}
\newblock
\APACjournalVolNumPages{Advances in neural information processing
  systems}{21}{}{}.
\PrintBackRefs{\CurrentBib}

\bibitem [\protect \citeauthoryear {%
Kestenbaum%
, Termine%
\BCBL {}\ \BBA {} Spelke%
}{%
Kestenbaum%
\ \protect \BOthers {.}}{%
{\protect \APACyear {1987}}%
}]{%
kestenbaum1987perception}
\APACinsertmetastar {%
kestenbaum1987perception}%
\begin{APACrefauthors}%
Kestenbaum, R.%
, Termine, N.%
\BCBL {}\ \BBA {} Spelke, E\BPBI S.%
\end{APACrefauthors}%
\unskip\
\newblock
\APACrefYearMonthDay{1987}{}{}.
\newblock
{\BBOQ}\APACrefatitle {Perception of objects and object boundaries by
  3-month-old infants} {Perception of objects and object boundaries by
  3-month-old infants}.{\BBCQ}
\newblock
\APACjournalVolNumPages{British journal of developmental
  psychology}{5}{4}{367--383}.
\PrintBackRefs{\CurrentBib}

\bibitem [\protect \citeauthoryear {%
Mao%
, Yang%
, Zhang%
, Goodman%
\BCBL {}\ \BBA {} Wu%
}{%
Mao%
\ \protect \BOthers {.}}{%
{\protect \APACyear {2022}}%
}]{%
maoclevrer}
\APACinsertmetastar {%
maoclevrer}%
\begin{APACrefauthors}%
Mao, J.%
, Yang, X.%
, Zhang, X.%
, Goodman, N.%
\BCBL {}\ \BBA {} Wu, J.%
\end{APACrefauthors}%
\unskip\
\newblock
\APACrefYearMonthDay{2022}{}{}.
\newblock
{\BBOQ}\APACrefatitle {CLEVRER-Humans: Describing Physical and Causal Events
  the Human Way} {Clevrer-humans: Describing physical and causal events the
  human way}.{\BBCQ}
\newblock
\BIn{} \APACrefbtitle {Thirty-sixth Conference on Neural Information Processing
  Systems Datasets and Benchmarks Track.} {Thirty-sixth conference on neural
  information processing systems datasets and benchmarks track.}
\PrintBackRefs{\CurrentBib}

\bibitem [\protect \citeauthoryear {%
Piloto%
, Weinstein%
, Battaglia%
\BCBL {}\ \BBA {} Botvinick%
}{%
Piloto%
\ \protect \BOthers {.}}{%
{\protect \APACyear {2022}}%
}]{%
piloto2022intuitive}
\APACinsertmetastar {%
piloto2022intuitive}%
\begin{APACrefauthors}%
Piloto, L\BPBI S.%
, Weinstein, A.%
, Battaglia, P.%
\BCBL {}\ \BBA {} Botvinick, M.%
\end{APACrefauthors}%
\unskip\
\newblock
\APACrefYearMonthDay{2022}{}{}.
\newblock
{\BBOQ}\APACrefatitle {Intuitive physics learning in a deep-learning model
  inspired by developmental psychology} {Intuitive physics learning in a
  deep-learning model inspired by developmental psychology}.{\BBCQ}
\newblock
\APACjournalVolNumPages{Nature human behaviour}{6}{9}{1257--1267}.
\PrintBackRefs{\CurrentBib}

\bibitem [\protect \citeauthoryear {%
Sch{\"o}lkopf%
\ \protect \BOthers {.}}{%
Sch{\"o}lkopf%
\ \protect \BOthers {.}}{%
{\protect \APACyear {2021}}%
}]{%
scholkopf2021toward}
\APACinsertmetastar {%
scholkopf2021toward}%
\begin{APACrefauthors}%
Sch{\"o}lkopf, B.%
, Locatello, F.%
, Bauer, S.%
, Ke, N\BPBI R.%
, Kalchbrenner, N.%
, Goyal, A.%
\BCBL {}\ \BBA {} Bengio, Y.%
\end{APACrefauthors}%
\unskip\
\newblock
\APACrefYearMonthDay{2021}{}{}.
\newblock
{\BBOQ}\APACrefatitle {Toward causal representation learning} {Toward causal
  representation learning}.{\BBCQ}
\newblock
\APACjournalVolNumPages{Proceedings of the IEEE}{109}{5}{612--634}.
\PrintBackRefs{\CurrentBib}

\bibitem [\protect \citeauthoryear {%
Sloman%
\ \BBA {} Lagnado%
}{%
Sloman%
\ \BBA {} Lagnado%
}{%
{\protect \APACyear {2004}}%
}]{%
sloman2004causal}
\APACinsertmetastar {%
sloman2004causal}%
\begin{APACrefauthors}%
Sloman, S.%
\BCBT {}\ \BBA {} Lagnado, D\BPBI A.%
\end{APACrefauthors}%
\unskip\
\newblock
\APACrefYearMonthDay{2004}{}{}.
\newblock
{\BBOQ}\APACrefatitle {Causal invariance in reasoning and learning} {Causal
  invariance in reasoning and learning}.{\BBCQ}
\newblock
\APACjournalVolNumPages{Psychology of learning and motivation}{44}{}{287--326}.
\PrintBackRefs{\CurrentBib}

\bibitem [\protect \citeauthoryear {%
Spelke%
}{%
Spelke%
}{%
{\protect \APACyear {1990}}%
}]{%
spelke1990principles}
\APACinsertmetastar {%
spelke1990principles}%
\begin{APACrefauthors}%
Spelke, E\BPBI S.%
\end{APACrefauthors}%
\unskip\
\newblock
\APACrefYearMonthDay{1990}{}{}.
\newblock
{\BBOQ}\APACrefatitle {Principles of object perception} {Principles of object
  perception}.{\BBCQ}
\newblock
\APACjournalVolNumPages{Cognitive science}{14}{1}{29--56}.
\PrintBackRefs{\CurrentBib}

\bibitem [\protect \citeauthoryear {%
Spelke%
}{%
Spelke%
}{%
{\protect \APACyear {2022}}%
}]{%
spelke2022babies}
\APACinsertmetastar {%
spelke2022babies}%
\begin{APACrefauthors}%
Spelke, E\BPBI S.%
\end{APACrefauthors}%
\unskip\
\newblock
\APACrefYear{2022}.
\newblock
\APACrefbtitle {What Babies Know: Core Knowledge and Composition Volume 1}
  {What babies know: Core knowledge and composition volume 1}\ (\BVOL~1).
\newblock
\APACaddressPublisher{}{Oxford University Press}.
\PrintBackRefs{\CurrentBib}

\bibitem [\protect \citeauthoryear {%
Spelke%
\ \BBA {} Kinzler%
}{%
Spelke%
\ \BBA {} Kinzler%
}{%
{\protect \APACyear {2007}}%
}]{%
spelke2007core}
\APACinsertmetastar {%
spelke2007core}%
\begin{APACrefauthors}%
Spelke, E\BPBI S.%
\BCBT {}\ \BBA {} Kinzler, K\BPBI D.%
\end{APACrefauthors}%
\unskip\
\newblock
\APACrefYearMonthDay{2007}{}{}.
\newblock
{\BBOQ}\APACrefatitle {Core knowledge} {Core knowledge}.{\BBCQ}
\newblock
\APACjournalVolNumPages{Developmental science}{10}{1}{89--96}.
\PrintBackRefs{\CurrentBib}

\bibitem [\protect \citeauthoryear {%
Spelke%
, von Hofsten%
\BCBL {}\ \BBA {} Kestenbaum%
}{%
Spelke%
\ \protect \BOthers {.}}{%
{\protect \APACyear {1989}}%
}]{%
spelke1989object}
\APACinsertmetastar {%
spelke1989object}%
\begin{APACrefauthors}%
Spelke, E\BPBI S.%
, von Hofsten, C.%
\BCBL {}\ \BBA {} Kestenbaum, R.%
\end{APACrefauthors}%
\unskip\
\newblock
\APACrefYearMonthDay{1989}{}{}.
\newblock
{\BBOQ}\APACrefatitle {Object perception in infancy: Interaction of spatial and
  kinetic information for object boundaries.} {Object perception in infancy:
  Interaction of spatial and kinetic information for object boundaries.}{\BBCQ}
\newblock
\APACjournalVolNumPages{Developmental Psychology}{25}{2}{185}.
\PrintBackRefs{\CurrentBib}

\bibitem [\protect \citeauthoryear {%
Tang%
\ \BBA {} Ellis%
}{%
Tang%
\ \BBA {} Ellis%
}{%
{\protect \APACyear {2022}}%
}]{%
tang2022perception}
\APACinsertmetastar {%
tang2022perception}%
\begin{APACrefauthors}%
Tang, H.%
\BCBT {}\ \BBA {} Ellis, K.%
\end{APACrefauthors}%
\unskip\
\newblock
\APACrefYearMonthDay{2022}{}{}.
\newblock
{\BBOQ}\APACrefatitle {From perception to programs: regularize,
  overparameterize, and amortize} {From perception to programs: regularize,
  overparameterize, and amortize}.{\BBCQ}
\newblock
\APACjournalVolNumPages{arXiv preprint arXiv:2206.05922}{}{}{}.
\PrintBackRefs{\CurrentBib}

\bibitem [\protect \citeauthoryear {%
Van Den~Oord%
, Vinyals%
\BCBL {}\ \protect \BOthers {.}}{%
Van Den~Oord%
\ \protect \BOthers {.}}{%
{\protect \APACyear {2017}}%
}]{%
van2017neural}
\APACinsertmetastar {%
van2017neural}%
\begin{APACrefauthors}%
Van Den~Oord, A.%
, Vinyals, O.%
\BCBL {}\ \BOthersPeriod {.}\end{APACrefauthors}%
\unskip\
\newblock
\APACrefYearMonthDay{2017}{}{}.
\newblock
{\BBOQ}\APACrefatitle {Neural discrete representation learning} {Neural
  discrete representation learning}.{\BBCQ}
\newblock
\APACjournalVolNumPages{Advances in neural information processing
  systems}{30}{}{}.
\PrintBackRefs{\CurrentBib}

\bibitem [\protect \citeauthoryear {%
Xu%
\ \BBA {} Carey%
}{%
Xu%
\ \BBA {} Carey%
}{%
{\protect \APACyear {1996}}%
}]{%
xu1996infants}
\APACinsertmetastar {%
xu1996infants}%
\begin{APACrefauthors}%
Xu, F.%
\BCBT {}\ \BBA {} Carey, S.%
\end{APACrefauthors}%
\unskip\
\newblock
\APACrefYearMonthDay{1996}{}{}.
\newblock
{\BBOQ}\APACrefatitle {Infants’ metaphysics: The case of numerical identity}
  {Infants’ metaphysics: The case of numerical identity}.{\BBCQ}
\newblock
\APACjournalVolNumPages{Cognitive psychology}{30}{2}{111--153}.
\PrintBackRefs{\CurrentBib}

\bibitem [\protect \citeauthoryear {%
Yang%
\ \BBA {} Piantadosi%
}{%
Yang%
\ \BBA {} Piantadosi%
}{%
{\protect \APACyear {2022}}%
}]{%
yang2022one}
\APACinsertmetastar {%
yang2022one}%
\begin{APACrefauthors}%
Yang, Y.%
\BCBT {}\ \BBA {} Piantadosi, S\BPBI T.%
\end{APACrefauthors}%
\unskip\
\newblock
\APACrefYearMonthDay{2022}{}{}.
\newblock
{\BBOQ}\APACrefatitle {One model for the learning of language} {One model for
  the learning of language}.{\BBCQ}
\newblock
\APACjournalVolNumPages{Proceedings of the National Academy of
  Sciences}{119}{5}{e2021865119}.
\PrintBackRefs{\CurrentBib}

\end{thebibliography}

\end{document}